\documentclass[a4paper,11pt]{article}
\usepackage{pos}
\usepackage{siunitx}
\usepackage{nicefrac}
\usepackage{booktabs}
\usepackage{csquotes}
\usepackage{float}
\usepackage{placeins}

\title{Three-dimensional Search for Annihilating Dark Matter in CBe dSph with the MAGIC Telescopes}
\ShortTitle{3D Search for Annihilating Dark Matter in CBe dSph with the MAGIC Telescopes}
\author*[a]{Stefan Fröse}
\author{Dominik Martin Elsässer\textsuperscript{\normalfont b} for the MAGIC Collaboration}
\author[c, d]{Hendrik Hildebrandt}
\author[c]{Elisa Pueschel}

\affiliation[a]{Institue of Physic, Academia Sinica,\\
  No. 128, Sec. 2, Academia Rd., Nangang Dist., Taipei City 115201, Taiwan (R.O.C.)}

\affiliation[b]{Department of Physics, TU Dortmund University,\\
Otto-Hahn-Straße 4a, 44227 Dortmund, Germany}
\affiliation[c]{Astronomical Institute (AIRUB), Faculty of Physics and Astronomy, Ruhr University Bochum, \\ 
Universitätsstr. 150, 44801 Bochum, Germany}
\affiliation[d]{German Centre for Cosmological Lensing (GCCL), Ruhr University Bochum, \\ 
Universitätsstr. 150, 44801 Bochum, Germany}

\emailAdd{stefanfroese@as.edu.tw}

\abstract{
  Dark matter (DM) candidates, such as Weakly Interacting Massive Particles (WIMPs), 
  can annihilate to Standard Model particles, subsequently producing gamma rays.
  In this work, we search for DM-induced gamma-ray signals from Coma Berenices dwarf spheroidal galaxy (CBe dSph)
  using approximately 25 hours of observations carried out by the Major Atmospheric Gamma Imaging Cherenkov (MAGIC) Telescope, 
  located at Roque de los Muchachos Observatory, La Palma, Spain.
  Building upon preceding analyses in the gamma-ray regime, we extend the DM search into three dimensions 
  by incorporating spatial information from the assumed DM-density distribution.
  This approach enhances sensitivity by leveraging both energy and spatial characteristics of the expected signal.
  The three-dimensional search for a faint signal necessitates the construction of a 
  background model, leading to the proposal of the \textit{exclusion-rotation} method. 
  This method stacks all observations, excludes the source region, and corrects for the 
  Azimuth-dependent acceptance of the telescopes by rotating the model.
  Furthermore, the open-source Python package \texttt{TITRATE} is presented,  
  introducing \textit{Asimov} datasets to the high-level analysis tool \texttt{Gammapy} for the
  approximation of the test statistic.
  No evidence of a DM-induced signal for annihilation to $b\bar{b}$, $W^+W^-$, $\mu^+\mu^-$, and $\tau^+\tau^-$ 
  and DM masses $m_\chi$ between \SI{0.17}{\TeV} and \SI{100}{\TeV} in CBe dSph is found. 
  Consequently, the first upper limits on the thermally averaged cross-section in energy and spatial dimensions 
  using \texttt{TITRATE} are set,
  leading to an improvement in sensitivity over previous results by MAGIC with respect to the assumed DM density in the target halo.  
  Moreover, the development of the Asimov datasets for DM search reduces the need for Monte Carlo simulations in estimating the mean and uncertainty bands of the sensitivity,
  paving the way for computationally efficient and scalable large-scale analyses across multiple targets and cosmic messengers.
}

\ConferenceLogo{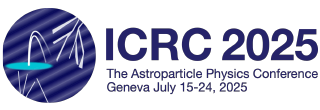}

\FullConference{39th International Cosmic Ray Conference (ICRC2025)\\
 15–24 July 2025\\
Geneva, Switzerland\\}
\begin{document}
\maketitle

\section{Introduction}
\noindent Dark Matter (DM) candidates, such as the Weakly Interacting Massive Particles (WIMPs) under investigation here, 
include variants that are capable of annihilating into Standard Model (SM) particles~\cite{Cirelli:2010xx,Slatyer:2017sev}.
Despite the suppression of direct photon production~\cite{Bergstrom:1997fh}, the subsequent production of high-energy gamma rays through alternative channels remains a possibility.
For astrophysical targets the expected gamma-ray flux is determined by two factors: firstly, the astrophysical component (AP), or J-factor in case of annihilating DM, and secondly, the particle physics component (PP).
The former represents the DM content within the target region along the line-of-sight, and the latter the primary interaction of DM.
DM-rich targets, such as dwarf spheroidal galaxies (dSphs) offer a promising pathway for the search for DM due to their high mass-to-light ratio~\cite{Charbonnier:2011ft}, 
leading to a relatively large AP contribution, and no or negligible gamma-ray background emission. 
\footnote{While other targets, such as the Galactic center, display more DM content compared to dSphs, the problem of separating the gamma-ray background and a potential DM signal 
remains a challenge among researches to date.}
Consequently, the Coma Berenices (CBe) dSph is considered, due to its proximity and low astrophysical background~\cite{Grcevich:2009gt, Geringer-Sameth:2014yza}.
Among the instruments capable of searching for DM-induced gamma-ray signals, such as satellites as Fermi-LAT~\cite{Fermi-LAT:2009ihh},   
Imaging Air Cherenkov Telescopes (IACTs) are capable of probing high-energy regions ($\mathcal{O}(\si{\TeV})$) of the PP contribution, 
by leveraging the in-air Cherenkov effect of gamma-ray-induced atmospheric showers, leading to an increased detection area over
alternative telescope designs.
To this end, the Major Atmospheric Gamma Imaging Cherenkov (MAGIC) Telescope system~\cite{MAGIC:2014zas} carried out a campaign of approximately $\SI{50}{\hour}$ of CBe dSph
observations, from which $\SI{25}{\hour}$ of high-quality data were selected for the analysis presented here. 
\noindent Previous MAGIC analyses on dSphs relied primarily on one-dimensional spectral fits~\cite{MAGIC:2021mog}, neglecting potential spatial information inherent in the expected DM signal distribution. 
Such simplified approaches may limit sensitivity, especially for faint gamma-ray signals from targets like the CBe dSph.
To enhance detection capabilities, this work introduces a three-dimensional search method that integrates spatial information derived from the assumed DM density profile.
However, employing spatially resolved data requires a robust method for precise background estimation.

\noindent We propose the \textit{exclusion-rotation} (ER) method to construct an accurate and azimuthally corrected background model for stacked observations.
Independent of the background method the newly developed \texttt{TITRATE}~\cite{stefan_frose_2025_14749249} Python package introduces Asimov datasets into the \texttt{Gammapy}~\cite{Gammapy:2023gvb} analysis framework, significantly 
reducing computational costs previously associated with extensive simulation of toy datasets for sensitivity estimation.
This contribution presents the first results of the three-dimensional DM search in the CBe dSph using MAGIC telescope data,
setting new constraints on the DM thermally averaged annihilation cross-section and comparing these new results to the one-dimensional approach.  

\section{Background Modelling}
\noindent The ER method estimates the background, defined as the observed events falsely classified as gamma rays, directly from the data. 
A three-dimensional histogram, in general referred to as a \textit{cube}, with 
the axes $\mathcal{A} = (\hat{E}, \hat{\text{FoV}}_\text{Az}, \hat{\text{FoV}}_\text{Alt})$ is defined as the background model. 
Here, $\hat{E}$ denotes the reconstructed energy and $\hat{\text{FoV}}_\text{Az}, \hat{\text{FoV}}_\text{Alt}$ the reconstructed direction in the Field-of-View (FoV) Alt-Az frame for the observed events.
The \textit{exclusion-only} method to populate the histogram is as follows:

\noindent Create three empty cubes $N_\text{eff}, T_\text{eff}, T_\text{tot}$ each with axes $\mathcal{A}$. Iterate over all observations and 
populate the $N_\text{eff}$ with the events of the observation while excluding all events within a
circular region of radius $r=\SI{0.4}{\deg}$\footnote{$r=\SI{0.4}{\deg}$ is equal to the MAGIC wobble distance~\cite{MAGIC:2014zas}} centered on the observation target.
$T_\text{eff}$ and $T_\text{tot}$ are filled simultaneously with the exposure time, with \textit{eff} denoting the cube with the source region excluded 
and \textit{tot} with it included.

\begin{figure}
  \begin{minipage}{0.48\textwidth}
    \centering
    \includegraphics[width=\textwidth]{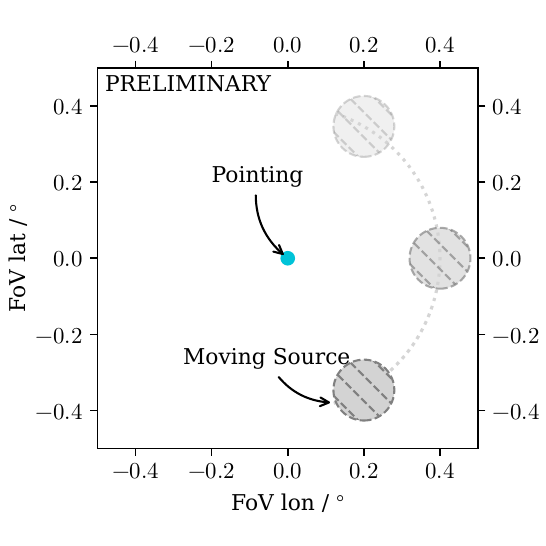}
    \caption{Schematic of the source rotation relative to the pointing position within the Alt-Az FoV during the observation.}
    \label{fig:rotation}
  \end{minipage}
  \hfill
  \begin{minipage}{0.51\textwidth}
    \centering
    \includegraphics[width=\textwidth]{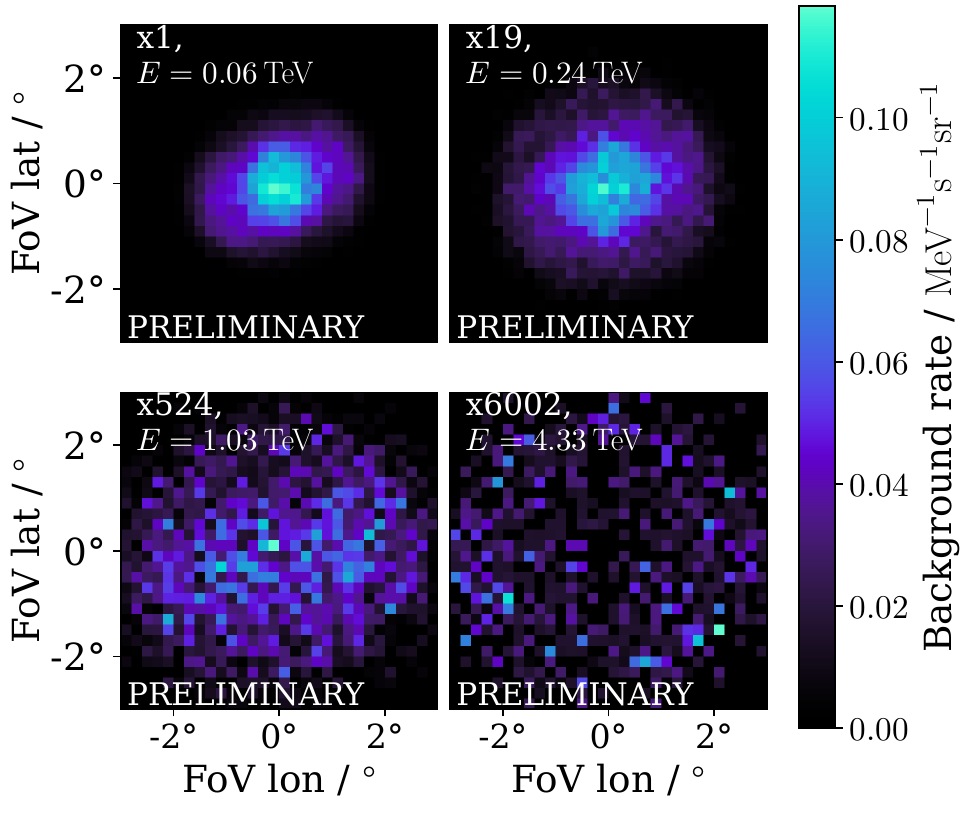}
    \caption{
      Background model of an arbitrary observation created with the novel \textit{exclusion-rotation} (ER) method.
      Each background map is multiplied by the factor denoted
      in the sub-figures to match the rate of the first energy bin.
    }\label{fig:bkg_model}
  \end{minipage}
\end{figure}

\noindent In practice, the exclusion region can be created in the average Alt-Az frame of each observation.
However, as \autoref{fig:rotation} illustrates, the rotation of the source in the FoV during an observation is not negligible.
To mitigate the impact of a falsely located exclusion region, the observations are divided into sub-observations with reduced live time.
The final background model is defined as 
  \begin{equation}
    \frac{\mathrm{d}N}{\mathrm{d}t\mathrm{d}E\mathrm{d}\Omega} = \frac{N_\text{eff}}{(\nicefrac{T_\text{eff}}T_\text{tot})\Delta\Omega\Delta E T_\text{obs}},
    \label{eq:bkg}
  \end{equation}
where $\Delta\Omega$ and $\Delta E$ define the solid angle and energy width of each bin in the cube, while $T_\text{obs}$ represents the total time of all observations.

\noindent According to Mender et al.~\cite{Mender:2023var}, the MAGIC telescopes exhibit  
a highly asymmetric \textit{stereo blob}\footnote{The stereo blob arises from the viewcone overlap of both telescopes.}. 
The empirical law between the orientation of the stereo blob, given by the rotation angle $\gamma_\text{theory}$,
and the Az alignment of the MAGIC view cones has been established as
\begin{equation}
  \gamma_\text{theory} = Az - \Delta\gamma,
  \label{eq:az_rotation}
\end{equation}
with the offset of $\Delta\gamma=\SI{34.23}{\degree}$.
Although a division of the background model in multiple Az bins is possible to compensate for the rotation of the acceptance among the observations, 
doing so leads to a decrease in statistics proportional to the number of bins, which therefore leads to a decrease of sensitivity.

\noindent We propose the novel ER method\footnote{\noindent An implementation of this method within the \texttt{BAccMod}~\cite{bony} Python package can be found on GitHub \url{https://github.com/mdebony/BAccMod/pull/40}}
 based on the preceding \text{exclusion} method, which de-rotates the events within one 
sub-observation according to the angle $\gamma_\text{PCA}$ from the first component of the Principal Component Analysis (PCA)~\cite{Shlens:2014uva} applied to the spatial event distribution 
in every energy bin.
The de-rotated events are re-rotated according to \autoref{eq:az_rotation} for the Az pointing of the observation. This procedure allows the use of all de-rotated events 
of all observations to construct one background model per Az pointing (one per observation). 
To avoid the influence of secondary effects of the zenith distance (Zd) on the background,
the method is implemented in three Zd bins: 5-15, 15-25, and 25-35\,\si{\deg}.
\noindent The background models are created in $30$ bins in the range $-\SI{3}{\degree}$ to $\SI{3}{\degree}$ for the Alt-Az and $8$ bins per decade in the range $\SI{0.05}{\TeV}$
to $\SI{5}{\TeV}$ for the energy axes.
\autoref{fig:bkg_model} displays the model of an arbitrary observation of the CBe dSph constructed with the ER method for four of sixteen
energy bins.
For low energies, the background is concentrated in the
center of the FoV showing the characteristic stereo blob. For increased energy, the background model
exhibits a distinctive “donut”-like shape emerging from the incapability of IACTs reconstructing showers 
with large Cherenkov footprints not fitting within the telescope camera. Reconstructable fully-enclosed shower images originate outside the central area
of the FoV.

%
% \begin{table}
%   \caption{Axial dimensions of the data cube used for background model creation.}\label{tab:3dmodel}
%   \begin{center}
%     \begin{tabular}[c]{cccc}
%       \toprule 
%       Axis & Range & Bins & per Decade\\
%       \midrule
%       Alt & \SIrange{-3}{3}{\degree}& 30 & no\\
%       Az & \SIrange{-3}{3}{\degree}& 30 & no\\
%       Energy & \SIrange{0.05}{5}{\TeV}& 8 & yes\\
%       \bottomrule
%     \end{tabular}
%   \end{center}
% \end{table}
%

\section{Asimov Datasets}
\noindent The calculation of the detection significance or upper limits of a DM model parameter on the observed data necessitates the computation of the  
p-value
% \begin{equation}
%   p = \int_{q_{\mu,\text{obs}}}^\infty f(q_\mu\vert\mu,\vec{\theta}) \mathrm{d}q_\mu
%   \label{eq:pval}
% \end{equation}
of the distribution $f(q_\mu\vert\mu,\vec{\theta})$ of the test statistic (TS) $q_\mu$, defined as the profiled log-likelihood ratio (pLLR).
Here, $\mu$ denotes the strength parameter of the signal and $\vec{\theta}$ the nuisance parameters. 
% The integration is bound by the TS evaluated on the measured dataset $q_{\mu,\text{obs}}$.
A common approximation for $f(q_\mu\vert\mu,\vec{\theta})$ is a $\chi^2$ distribution, according to Wilks' theorem~\cite{Wilks:1938dza}.
However, two main problems arise from using the Wilks approximation: firstly, if the estimator $\hat{\mu}$
\footnote{From here on all quantities denoted as $\hat{.}$ represent estimators on the respective quantity.} is smaller than zero after 
calculating the pLLR, the statistical test loses physical meaning and secondly, a vast amount of computationally costly toy Monte Carlo (MC) datasets is necessary
to compute the background-only expectation (sensitivity). 

\noindent We propose the usage of the \emph{Asimov datasets} published by Cowan et al.~\cite{Cowan:2010js}
to approximate the TS distribution without relying on extensive MC simulations. 
The Asimov datasets are constructed such that the estimators 
fulfill $\hat{\mu} = \mu^\prime$ and $\hat{\vec{\theta}} = \vec{\theta}^\prime$, 
representing the expected outcome under the hypothesis $(\mu^\prime, \vec{\theta}^\prime)$.

\noindent To ensure valid statistical inference for the unphysical case, $\hat{\mu} < 0$, 
we employ the \emph{modified test statistic} $\tilde{q}_\mu$, 
which remains well-defined at the boundary. 
The approximation of the TS distribution as a non-central $\chi^2$ allows for the use of the $\mathrm{CL}_\mathrm{s}$\footnote{Although this denotes the name of the method, it is 
notable that CL is short for confidence level. See~\cite{Read:2002hq} for more details.} method~\cite{Read:2002hq} for the first time in a MAGIC analysis,
yielding the computation of upper limits and expected sensitivity bands $\pm N\sigma$~\cite{Gross:2018okg}:
\begin{equation}
p_\mu \overset{!}{=} \frac{1-\Phi(\sqrt{\tilde{q}_\mu})}{\Phi(\sqrt{\tilde{q}_{\mu, A}}-\sqrt{\tilde{q}_\mu})}\,,
\end{equation}
\begin{equation}
  \mu_{\text{exp}} = \sigma(\Phi^{-1}(1 - \alpha\Phi(N)) \pm N) \,,
  \label{eq:null}
\end{equation}
at confidence level $1-\alpha$, while $\Phi(.)$ denotes the cumulative density function (CDF) of the standard normal distribution and index \enquote{A} denotes the evaluation of the TS on the Asimov dataset.
We provide an implementation of the Asimov-based computations through the newly developed Python package \texttt{TITRATE}\footnote{\url{https://github.com/StFroese/TITRATE}}.

\section{Results}
\begin{figure}[t]
  \begin{center}
    \includegraphics[width=\textwidth]{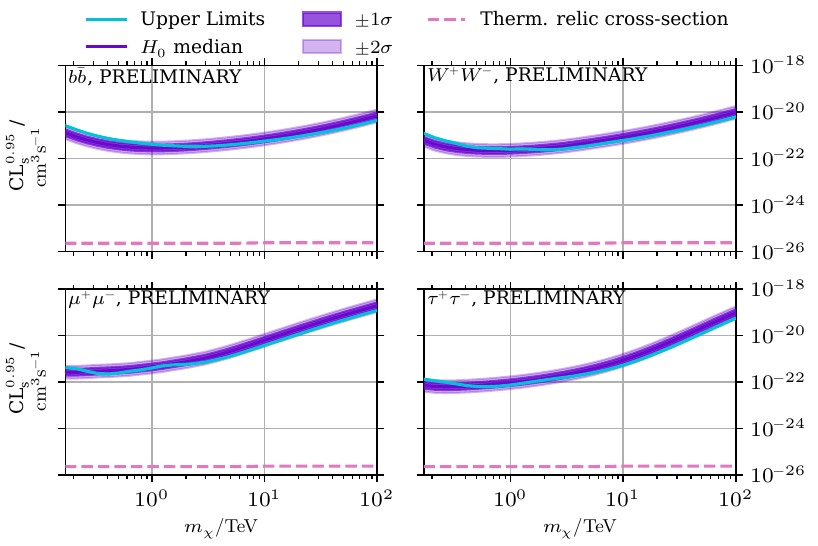}
  \end{center}
  \caption{
    Upper limits and sensitivity bands on the thermally averaged cross-section for the 3D analysis. Thermal relic cross-section taken from~\cite{Steigman:2012nb}.
  }\label{fig:uls}
\end{figure}

\begin{figure}
  \begin{center}
    \includegraphics[width=\textwidth]{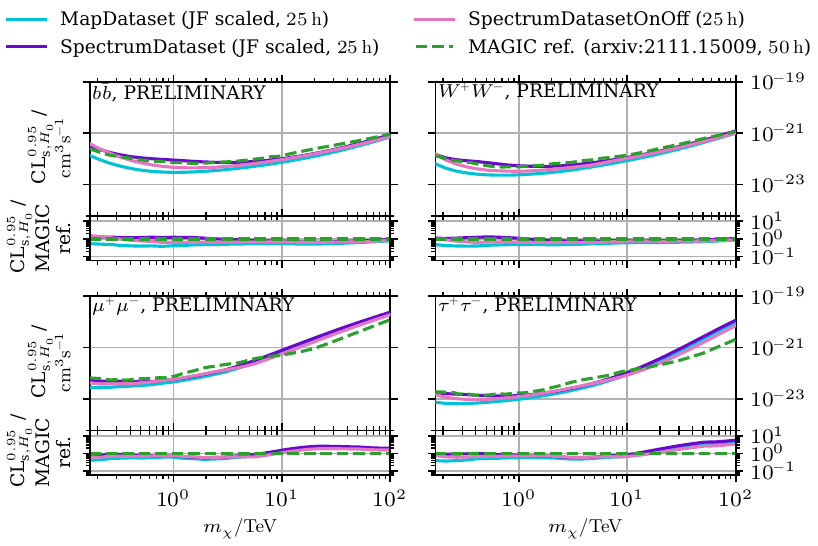}
  \end{center}
  \caption{
    Median expected limits on the thermally averaged cross-section for the \texttt{Gammapy} datasets and the preceding MAGIC analysis~\cite{MAGIC:2021mog}, as well as their ratios.
  }\label{fig:ratios}
\end{figure}
\setlength{\textfloatsep}{6pt plus 1pt minus 1pt}  % default is ~20pt
\noindent No significant excess was observed in approximately $\SI{25}{\hour}$ of MAGIC observations of the CBe dSph. We compute 95\% 
$\mathrm{CL}_\mathrm{s}$ upper limits on the thermally averaged annihilation cross-section as a function of DM mass $m_\chi \in [0.17, 100] \si{TeV}$
for four annihilation channels: $b\bar{b}$, $W^+W^-$, $\mu^+\mu^-$, and $\tau^+\tau^-$. The empirical modeling of the AP contribution is taken from Geringer-Sameth et al.~\cite{Geringer-Sameth:2014yza}.\\
\noindent \autoref{fig:uls} shows the limits derived from the 3D analysis using \texttt{TITRATE} and Asimov datasets. The median expected limits ($H_0$) 
and $\pm 1\sigma$, $\pm 2\sigma$ containment bands are computed with~\autoref{eq:null}. The observed limits lie within the bands,
consistent with background-only expectation. Additionally, the thermal relic cross-section taken from Steigman et al.~\cite{Steigman:2012nb} is displayed.
To evaluate the performance of the new analysis method, the expected limits are
obtained with the \texttt{Gammapy}~\cite{Gammapy:2023gvb} \texttt{MapDataset} (3D, spatial+energy), \texttt{SpectrumDataset} (1D, energy-only), and \texttt{SpectrumDatasetOnOff} (1D, energy-only).
While the \texttt{SpectrumDataset} is constructed via integration of the spatial axes of the 
\texttt{MapDataset} in a region of $\theta^2=\SI{0.03}{\deg \squared}$\footnote{This size is given by the ON-region size of the preceding MAGIC publication~\cite{MAGIC:2021mog}} around the source position, 
the \texttt{SpectrumDatasetOnOff} does not rely on the ER method presented, but rather constructs the background from Off-regions.
\autoref{fig:ratios} displays a comparison of the sensitivities to the preceding MAGIC publication on the CBe dSph found in~\cite{MAGIC:2021mog}. 
It is notable, that the previously published limits rely on full observation time of the CBe dSph campaign, while 
this analysis relies on the high-quality $\SI{25}{\hour}$ set.
The 3D analysis exhibits an increase in sensitivity compared to the \texttt{SpectrumDataset}-based analysis in all channels and DM masses. 
While the 3D analysis shows the most sensitive limits for the non-leptonic channels and up to $\approx\SI{4}{\TeV}$ and $\approx\SI{7}{\TeV}$ for the leptonic channels among the \texttt{Gammapy} datasets, 
the \texttt{SpectrumDatasetOnOff}-based analysis 
is the most sensitive approach at higher mass ranges for the leptonic channels.
Overall, the 3D analysis illustrates improved limits for the non-leptonic channels and the leptonic channels up to $\approx\SI{10}{\TeV}$ 
with half the amount of exposure time in comparison to the previously published MAGIC results.
\vspace{-2ex}
\section{Summary}
\vspace{-2ex}
\noindent In this work, we have presented a three-dimensional search for gamma rays from DM annihilation in the CBe dSph
dwarf spheroidal galaxy using approximately $\SI{25}{\hour}$ of data from the MAGIC telescopes. 
We have introduce the ER method for background estimation and modelling.
This approach corrects for azimuth-dependent acceptance variations without relying on external simulations.
To estimate sensitivities without MC trials, we provide \texttt{TITRATE}, a Python package that integrates
Asimov dataset generation into \texttt{Gammapy}. 
This enables fast computation of the TS distribution and associated upper limits, as well as the introduction 
of the $\mathrm{CL}_\mathrm{s}$ method for gamma-ray analyses within Gammapy.
\texttt{TITRATE} further allows reproducible and scalable analysis across targets, 
making it suitable for future multi-messenger DM searches.
No excess consistent with DM annihilation into SM products was observed. We have set upper limits on the thermally
averaged cross-section for annihilation channels into $b\bar{b}$, $W^+W^-$, $\mu^+\mu^-$, and $\tau^+\tau^-$ in 
the mass range from \SI{0.17}{\TeV} to \SI{100}{\TeV}. 
By incorporating spatial information 
from the assumed DM distribution, the analysis gains a mass- and channel-dependent sensitivity improvement over traditional spectral-only approaches
with half the amount of exposure time.

\acknowledgments
\vspace{-2ex}
\noindent We would like to thank the Instituto de Astrof\'{\i}sica de Canarias for the excellent working conditions at the Observatorio del Roque de los Muchachos in La Palma. The financial support of the German BMBF, MPG and HGF; the Italian INFN and INAF; the Swiss National Fund SNF; the grants PID2019-107988GB-C22, PID2022-136828NB-C41, PID2022-137810NB-C22, PID2022-138172NB-C41, PID2022-138172NB-C42, PID2022-138172NB-C43, PID2022-139117NB-C41, PID2022-139117NB-C42, PID2022-139117NB-C43, PID2022-139117NB-C44, CNS2023-144504 funded by the Spanish MCIN/AEI/ 10.13039/501100011033 and "ERDF A way of making Europe; the Indian Department of Atomic Energy; the Japanese ICRR, the University of Tokyo, JSPS, and MEXT; the Bulgarian Ministry of Education and Science, National RI Roadmap Project DO1-400/18.12.2020 and the Academy of Finland grant nr. 320045 is gratefully acknowledged. This work was also been supported by Centros de Excelencia ``Severo Ochoa'' y Unidades ``Mar\'{\i}a de Maeztu'' program of the Spanish MCIN/AEI/ 10.13039/501100011033 (CEX2019-000920-S, CEX2019-000918-M, CEX2021-001131-S) and by the CERCA institution and grants 2021SGR00426 and 2021SGR00773 of the Generalitat de Catalunya; by the Croatian Science Foundation (HrZZ) Project IP-2022-10-4595 and the University of Rijeka Project uniri-prirod-18-48; by the Deutsche Forschungsgemeinschaft (SFB1491) and by the Lamarr-Institute for Machine Learning and Artificial Intelligence; by the Polish Ministry Of Education and Science grant No. 2021/WK/08; and by the Brazilian MCTIC, CNPq and FAPERJ.
H. Hildebrandt is supported by a DFG Heisenberg grant (Hi 1495/5-1), the DFG Collaborative Research Center SFB1491, an ERC Consolidator Grant (No. 770935), and the DLR project 50QE2305.
E. Pueschel is supported by an ERC Consolidator Grant (No. 101124914).

\bibliography{bibliography.bib}

\end{document}